\documentclass[aps,pra,twoside,twocolumn,superscriptaddress,floatfix,10pt]{revtex4-2}
\usepackage{etoolbox}
\newtoggle{Wiley}
\settoggle{Wiley}{false}

\usepackage{preamble}

\makeatletter
\newcommand{\mycolumnwidth}{\columnwidth}
\newcommand*{\addFileDependency}[1]{
  \typeout{(#1)}
  \@addtofilelist{#1}
  \IfFileExists{#1}{}{\typeout{No file #1.}}
}
\makeatother

\newcommand*{\myexternaldocument}[1]{%
    \externaldocument[SI-]{#1}%
    \addFileDependency{#1.tex}%
    \addFileDependency{#1.aux}%
}

\myexternaldocument{supp}

\begin{document}


\title{The Cycling Mechanism of Manganese-Oxide Cathodes in Zinc Batteries: A Theory-Based Approach}

\iftoggle{Wiley}
{
	\maketitle
}
{}

\iftoggle{Wiley}{
	\author{Niklas J. Herrmann}
	\author{Holger Euchner}
	\author{Axel Groß}
	\author{Birger Horstmann*}

	\begin{affiliations}
		N.J. Herrmann, Prof. Dr. A.Groß, Prof. Dr. B. Horstmann\\
		Helmholtz Institute Ulm (HIU)\\
		Helmholtzstr. 11, D-89081 Ulm, Germany\\
		Email Address: birger.horstmann@dlr.de\\

		N.J. Herrmann, Prof. Dr. B. Horstmann\\
		German Aerospace Center (DLR)\\
		Institute of Engineering Thermodynamics\\
		Wilhelm-Runge-Strasse 10, D-89081 Ulm, Germany\\

		Prof. Dr. B. Horstmann, Prof. Dr. A. Groß\\
		Institute of Electrochemistry\\
		Ulm University (UUlm)\\
		Albert-Einstein-Allee 47, D-89081 Ulm, Germany\\

		Dr. H. Euchner\\
		Institute of Physical and Theoretical Chemistry\\
		University of Tübingen\\
		Auf der Morgenstelle 18, D-72076 Tübingen, Germany\\
		
	\end{affiliations}
}
{
	\author{Niklas J. Herrmann}
	\affiliation{German Aerospace Center, Wilhelm-Runge-Straße 10, D-89081 Ulm, Germany}
	\affiliation{Helmholtz Institute Ulm, Helmholtzstra{\ss}e 11, D-89081 Ulm,
		Germany}  
	\affiliation{Universit\"at Ulm,
		Albert-Einstein-Allee 47, D-89081 Ulm, Germany} 
	
	\author{Holger Euchner}
	\affiliation{Universit\"at Ulm,
		Albert-Einstein-Allee 47, D-89081 Ulm, Germany}
	\affiliation{ Institute of Physical and Theoretical Chemistry, University of T\"ubingen,
		Auf der Morgenstelle 15, D-72076 Tübingen, Germany}
	
	\author{Axel Groß}
	\affiliation{Helmholtz Institute Ulm, Helmholtzstra{\ss}e 11, D-89081 Ulm, Germany}  
	\affiliation{Universit\"at Ulm,
		Albert-Einstein-Allee 47, D-89081 Ulm, Germany}

	\author{Birger Horstmann}
	\email{birger.horstmann@dlr.de}
	\affiliation{German Aerospace Center, Wilhelm-Runge-Straße 10, D-89081 Ulm, Germany}
	\affiliation{Helmholtz Institute Ulm, Helmholtzstra{\ss}e 11, D-89081 Ulm,
		Germany}  
	\affiliation{Universit\"at Ulm,
		Albert-Einstein-Allee 47, D-89081 Ulm, Germany} 
}


\begin{abstract}
Zinc-based batteries offer good volumetric energy densities and are compatible with environmentally friendly aqueous electrolytes.
Zinc-ion batteries (ZIBs) rely on a lithium-ion-like \ce{Zn^2+}-shuttle, which enables higher roundtrip efficiencies and better cycle life than zinc-air batteries.
Manganese-oxide cathodes in near-neutral zinc sulfate electrolytes are the most prominent candidates for ZIBs.
\ce{Zn^2+}-insertion, \ce{H+}-insertion, and \ce{Mn^2+}-dissolution are proposed to contribute to the charge-storage mechanism.
During discharge and charge, two distinct phases are observed.
Notably, the pH-driven precipitation of zinc-sulfate-hydroxide is detected during the second discharge phase.
However, a complete and consistent understanding of the two-phase mechanism of these ZIBs is still missing.
This paper presents a continuum full cell model supported by DFT calculations to investigate the implications of these observations.
We integrate the complex-formation reactions of near-neutral aqueous electrolytes into the battery model and,
in combination with the DFT calculations, draw a consistent picture of the cycling mechanism.
We investigate the interplay between electrolyte pH and reaction mechanisms at the manganese-oxide cathodes and identify the dominant charge-storage mechanism.
Our model is validated with electrochemical cycling data, cyclic voltammograms, and in-situ pH measurments. This allows us to analyse the influence of cell design and electrolyte composition on cycling and optimize the battery performance.
\end{abstract}

\iftoggle{Wiley}
{
}
{
	\maketitle
}

\allowdisplaybreaks
\section{Introduction}

Zinc-metal anodes feature competitive energy densities and are sufficiently stable in aqueous electrolytes, which are environmentally friendly, cheap and have excellent ionic conductivity.
Several primary zinc batteries based on alkaline electrolytes, such as zinc-carbon, zinc-air, or alkaline \ce{MnO2} batteries are commercially used for a long time \cite{Borchers2021}.
However, these alkaline zinc batteries were never successfully commercialized as secondary batteries as they experience very limited rechargeability \cite{Wruck1991,Kordesch1994}.

Modern rechargeable zinc-ion batteries use similar materials and electrodes as Zn-\ce{MnO2} batteries but with non-alkaline electrolytes \cite{Blanc2020}.
In 1986, Yamamoto and coworkers presented a battery with a metallic zinc anode and \ce{MnO2} cathode  \cite{Yamamoto1986, Shoji1988}.
In exchange for the KOH electrolyte, they used a near-neutral aqueous solution of \ce{ZnSO4} as electrolyte.
This early experiment showed a rechargeability, significantly better than their alkaline predecessors, but still limited to around 30 cycles \cite{Yamamoto1986}.
Different inorganic zinc salts were tested as aqueous electrolytes \cite{Shoji1988}, of which \ce{ZnSO4}, which is still the most popular \cite{Liu2021}, showed the highest achievable capacity. 
At the beginning of the 21\textsuperscript{st} century, improvements in cycling stability re-sparked interest in ZIBs, leading to a rapidly growing amount of research in the last decade. 
Several other cathode materials were tested \cite{Borchers2021}, vanadates
achieving high stabilities \cite{Liu2019,Wang2019c}, Prussian blue analogs with extraordinary cycling stability \cite{Trocoli2015}, and organic cathode materials with promising capacities \cite{Kundu2018, pop00145}. 
Nevertheless, manganese-based cathodes are still the most promising combining a well-established production chain with a competitive overall cell performance. 
There are several approaches to increase the cell voltage which typically requires extending the electrolyte stability by using non-aqueous electrolytes \cite{Kundu2018b, Corpuz2019}. 
However, aqueous electrolytes achieve higher energy densities and offer a price advantage and excellent eco-friendliness.

In the last decade, research achieved a significant increase in cycling stability and investigated the details of cycling characteristics. 
It was observed that both discharging as well as charging voltages show two distinct phases \cite{Wu2018, Godeffroy2022, Chen2022a, Chen2022}.
While the discharge and charge phases at high state of charge (SOC) experience a rather fast kinetic, the second phase at the end of discharge and the beginning of charge only shows slow kinetics \cite{Wu2018, Jaikrajang2021, Pu2022}.
A dip in cell potential is present between the two discharge phases. It is most pronounced in micro-structured \ce{MnO2} cathodes.
This voltage dip during discharge correlates with the onset of \ce{Zn4(OH)6SO4} (ZHS) precipitation.
Precipitation of ZHS occurs at the \ce{MnO2} cathode during the second phase of the discharge and is dissolved again at the beginning of charge as demonstrated by in-situ spectroscopy \cite{Putro2020,Chen2022}.
Additionally, different polymorphs of \ce{MnO2} are considered and studied as electrodes, but \textdelta-\ce{MnO2} with its layered structure is often regarded as the most promising \cite{Kim1998,Alfaruqi2015c,Jiang2020}.
Furthermore, the cell is optimized by varying electrolyte concentration and composition \cite{Liu2021, Wang2022}. 
Especially, pre-adding a \ce{Mn^2+}-salt to optimize cycling performance is evaluated \cite{Chamoun2018a, Soundharrajan2020, Park2022}.

The precipitation of ZHS indicates a reaction process that changes electrolyte pH. 
ZHS precipitates at a pH $\approx 5.5$, which is more alkaline than benign \ce{ZnSO4} electrolytes \cite{Lee2016a}.
The coinsertion of \ce{H+} is often attributed to the observed pH shift.
Lately, research papers focused on the dissolution process of \ce{Mn^2+}-ion leaching from the cathode \cite{Shi2021, Chamoun2018a, Guo2020}. 
Both, the insertion of \ce{H+} as well as the dissolution of \ce{Mn^2+} result in an increase in electrolyte pH \cite{Lee2016a, Fitz2021}.
Experiments with analytical measurements during cycling have shown that reversible variations of the \ce{Mn^2+} concentration in the electrolyte occur \cite{Wu2018, Chen2022a}. 
The importance of cathodic dissolution for understanding the cycling mechanism of \ce{MnO2}-based ZIBs is further highlighted by the recently published works of Chen et al. \cite{Chen2022}, Godeffroy et al. \cite{Godeffroy2022}, and Yang et al. \cite{Chen2022a}.

\begin{figure}[t]
	\includegraphics[width=\mycolumnwidth]{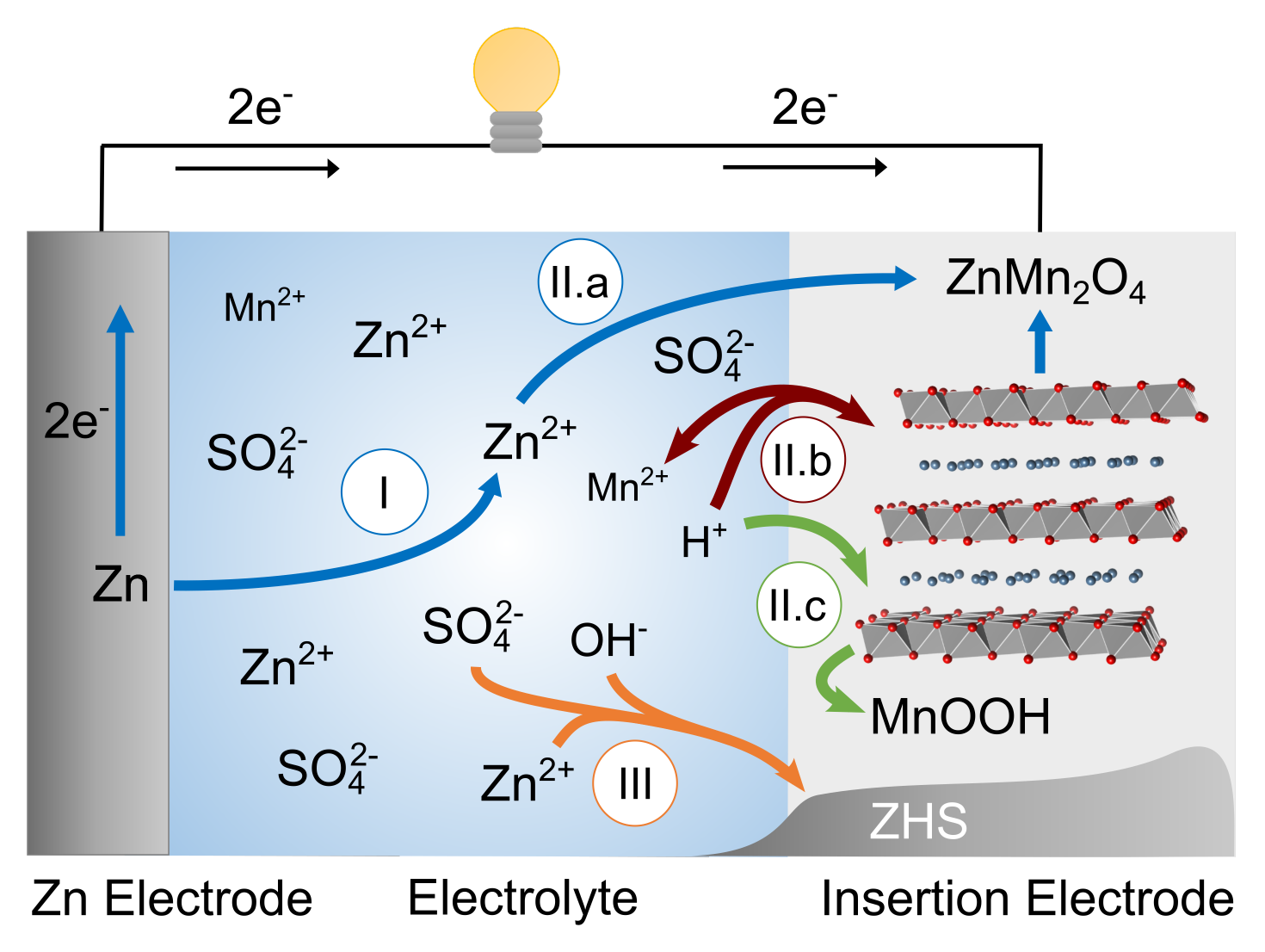}
	\caption{Schematic overview on proposed charge-storage mechanisms in ZIB. 
		The redox reaction of the zinc metal anode is shown on the left (I). At 
		the cathode, the electrochemical reactions are the \ce{Zn^2+}-Insertion (II.a), the \ce{Mn^2+} dissolution (II.b) and the insertion of \ce{H+} (II.c). The precipitation of ZHS (III), which is experimentally observed at the cathode, is shown in the lower right of the figure.}
	\label{fig:Schematic}
\end{figure}

In this paper, we present a theory-based approach and identify the cycling mechanism of ZIBs (see \cref{fig:Schematic}).
We focus on the behavior of ZIBs with \ce{MnO2} cathode in an aqueous \ce{ZnSO4} solution.
With the help of density functional theory (DFT) calculations, we investigate the properties of \textdelta-\ce{MnO2} and evaluate the dissolution and insertion potentials of the experimentally proposed processes \cite{Blanc2020}.
Additionally, we use thermodynamic calculations of the equilibrium speciation of the \ce{ZnSO4} electrolyte to identify electrolyte stability with respect to precipitation and to quantify the pH buffering properties. 
Based on this result, we develop our ZIB model describing the dynamic cell behavior. 
We implement a pseudo-two-dimensional (P2D) cell model which uses the quasi-particle transport theory derived by Clark et al. \cite{Clark2017, Clark2019, Clark2020b}.
We both investigate the zinc- and proton-insertion mechanism as well as \ce{Mn^2+}-dissolution (\cref{fig:Schematic}) and compare them with evidence from electrochemical cycling measurements. 
With this model, we elucidate the cycling mechanism of ZIB cells and use it for cell optimizations.

\section{Theory}

\subsection{Density Functional Theory (DFT)}
\label{ssec:theory_dft}
Density functional theory (DFT) is the standard tool for material simulations \cite{Liu2020b,Euchner2022}. Based on the \ce{MnO2} structure, we calculate the open circuit voltage (OCV) and compare different proposed reaction processes. 
For this purpose, we simulate the electronic structure of \ce{H_xZn_yMnO2 * H2O} with \ce{H} content $x \in [0,1]$ as well as \ce{Zn} content $y \in [0,0.5]$ 
and calculate the total energy $E_\mathrm{tot}$ of the relevant \ce{MnO2} structures for \ce{H+} and \ce{Zn^2+} insertion.
We approximate the overall difference in the Gibbs free energy $\Delta G$ as
\begin{equation}
	\Delta G \approx \Delta E - T \Delta S^{\mathrm{conf}}\ \mathrm{,}
\end{equation}
where $S^\mathrm{conf}$ is the configurational entropy of the structure and $\Delta E$ the difference in the total energies calculated by DFT. Extending the computational hydrogen electrode to the computational zinc electrode \cite{Liu2020b, Euchner2022}, we derive convenient expressions for the electrochemical potentials $\tilde{\mu}_i = \mu_i + z_ieU$ for \ce{Zn^2+} and \ce{H+} thus avoiding explicit calculations of solvation energies. 
This approach uses the circumstance, that the equilibria at standard conditions can be used to express the electrochemical potentials of solvated ions through molecular or atomic chemical potentials \cite{Liu2020b}. 
In detail, the definition for the standard hydrogen potential $U_\mathrm{SHE}$ uses the equilibrium of dissolved protons and hydrogen in the gas phase,  
\begin{align}
	\Delta \tilde{\mu}_{\ce{H+}} &= \tilde{\mu}_{\ce{H+(aq)}} + \tilde{\mu}_{e^-} - \frac{1}{2}E_{\ce{H2}} \nonumber \\
	&= - e U_{\mathrm{SHE}} - k_\mathrm{B} T \ln(10)\mathrm{pH}\ \mathrm{.}
	\label{eq:chempotH}
\end{align}
Analogous, the electrochemical potential for \ce{Zn^2+} in solution is calculated as
\begin{align}
	\Delta \tilde{\mu}_{\ce{Zn^{2+}}} &= \tilde{\mu}_{\ce{Zn^{2+}(aq)}} + 2\tilde{\mu}_{\ce{e}^-}  - E_{\ce{Zn}} \nonumber \\ 
	&= - 2e\left( U_{\mathrm{SHE}} - U_0\right)  
	- k_\mathrm{B} T \ln(a_{\ce{Zn^{2+}}})\ \mathrm{,}
	\label{eq:chempotZn}
\end{align}
where $U_0$ is the standard potential of zinc vs. $\mathrm{SHE}$.

Finally, we derive the insertion potential as 
$U_{\mathrm{ins}} = - \nicefrac{\Delta G}{z_i \Delta \mathrm{N}_i}$, where
\begin{align}
	\Delta G =&\ E_\mathrm{tot}^{\ce{H_xZn_yMnO2}} - E_\mathrm{tot}^{\ce{MnO2}} - T \cdot \left(S_\mathrm{conf}^{\ce{H_xZn_yMnO2}} - S_\mathrm{conf}^{\ce{MnO2}}\right)\nonumber \\
	\quad& - x \cdot \left(E_\mathrm{tot}^{{\mathrm{Zn(bulk)}}} + \Delta \tilde{\mu}_{\ce{Zn^2+}}\right) 
	- y \cdot \left(\frac{1}{2} E_\mathrm{tot}^\mathrm{H2(gas)} 
	+\Delta \tilde{\mu}_{\ce{H+}}\right)\ \mathrm{.} 
	\label{eq:deltaG_DFT}
\end{align}
The quantitative contribution of the configurational entropy can be found in the Supporting Information (\cref{SI-fig:configentropy}). At room-temperature, $T=\SI{300}{\kelvin}$, the relative influence is in the order of $1 k_\mathrm{B}T \approx \SI{25}{\milli\electronvolt}$.

Analogous reasoning leads to the dissolution potential $U_\mathrm{diss}$
of the respective \ce{H_xZn_yMnO2 * H2O} species. 
Here, we extend the calculations of the chemical potential to include \ce{Mn^2+}. The dissolution potential can then be calculated as $U_\mathrm{diss}  = - \nicefrac{\Delta G}{z_i \Delta \mathrm{N}_i}$.
For the dissolution, $\Delta G$ only includes the Gibbs free energy of the dissolved structure. 
All reaction products are dissolved ions, and their energy contributions are therefore included by their electrochemical potentials.
The full details can be found in the Supporting Information.

\subsection{Continuum Cell Model}
\subsubsection*{Equilibrium Speciation \& Quasi-Particle Transport model}
\label{ssec:qp}
\label{ssec:transport}

When simulating the transport of near-neutral aqueous electrolytes, we must follow the dynamics of multiple species formed by zinc and its ligands.
Not only is this computationally costly by increasing the problem dimensionality, but it also decreases numerical stability due to the different timescales and the non-linearity of complex-forming reaction kinetics.
Our approach builds upon the quasi-particle transport model developed and applied in previous works on near-neutral zinc-air batteries \cite{Clark2017, Clark2019, Clark2020b}. 

The presented quasi-particle framework utilizes an abstraction level to resolve the dynamic behavior in aqueous electrolytes. 
We define quasi-particles so that their concentrations are invariant under the complex-formation reactions. 
This allows us to decouple slow electrolyte transport and slow heterogeneous reactions from fast complex formation reactions. 
We calculate the transport of \ce{Zn^2+}-quasi-particles instead of each \ce{Zn}-ligand complex individually. 
On the side, we solve for electrolyte equilibrium speciation with algebraic equations defining the respective quasi-particles.
In this work, we use the quasi-particles $\ce{Zn^2+}_\mathrm{T}$, $\ce{H+}_\mathrm{T}$, $\ce{Mn^2+}_\mathrm{T}$ and $\ce{SO4^2-}_\mathrm{T}$.
The index $_\mathrm{T}$ denotes total concentration. 
For example, $\ce{Zn^2+}_\mathrm{T}$ is the total concentration of \ce{Zn} atoms, defined as
\begin{align}
	[\ce{Zn^2+}_\mathrm{T}] =& [\ce{Zn^2+}] + \sum_{n=1}^4 [\ce{Zn(SO4)_{n}^{2\cdot(1-n)}}] \nonumber \\
	\quad +& \sum_{n=1}^4 [\ce{Zn(OH)_{n}^{2-n}}] \nonumber \\
	\quad +& 2\cdot\left([\ce{Zn2OH^3+}] + [\ce{Zn2(OH)_6^2-}] \right) \nonumber \\
	\quad +& 4 \cdot [\ce{Zn4(OH)4^4+}]\ \mathrm{.}\label{eq:quasi-particle}
\end{align}
Here, $n$ is the stoichiometry of the zinc-sulfate complex, square brackets are used to indicate a concentration ($[\mathrm{X}] = c_\mathrm{X}$).
Consequently, electrolyte pH is given by the \ce{H+}-concentration as $\mathrm{pH} = - \log_{10} c_{\ce{H+}}/c_0$. We equate concentrations and activities as we analyzed that measured activity coefficients do not significantly alter our results.
All details of the quasi-particle Ansatz can be found in Ref. \citenum{Clark2019}. Our definitions of quasi-particles are given
in the Supporting Information (\cref{SI-ssec:electrolytethermo}). 

Homogeneous reactions govern the formation of complexes in the electrolyte.
In equilibrium, the law of mass action determines the ratio of reaction products and reactants. 
For example for \ce{Zn(SO4)2^2-}, the law of mass action reads
\begin{equation}
	\frac{c_{\ce{Zn(SO4)2^2-}}}{c_{\ce{Zn^2+}} ~ c_{\ce{SO4^2-}}^{2}}= \beta \ \mathrm{,}
\end{equation}
with $\beta = 10^{-3.28}$ from Ref. \citenum{Ball1991}.  
We use laws of mass action to express the concentrations on the right side of \cref{eq:quasi-particle} with the concentrations of the elementary ions $\ce{Zn^2+}$, $\ce{H+}$, $\ce{Mn^2+}$ and $\ce{SO4^2-}$.
By combining the resulting set of algebraic equations with the charge-neutrality equation for quasi-particles,
\begin{equation}
	0 =   2\cdot[\ce{Zn^2+}_\mathrm{T}] + [\ce{H+}_\mathrm{T}] + 2\cdot [\ce{Mn^2+}_\mathrm{T}] - 2\cdot [\ce{SO4^2-}_\mathrm{T}]\ \mathrm{,}
\end{equation} 
we calculate the concentrations of all complexes, i.e., the equilibrium electrolyte speciation.
The homogeneous electrolyte reactions for this work and the used stability constants \cite{Nordstrom1991,Ball1991,Baes1976,Grenthe2004} are given in the Supporting Information (\cref{SI-tab:electrolytespeciation}). 

We simulate transport for $4$ quasi-particles instead of $24$ complexes. Derived from consistent transport theory \cite{Stamm2017, Clark2019}, quasi-particle dynamics is calculated with the continuity equation
\begin{equation}
	\frac{\partial \epsilon_\mathrm{e} c_q}{\partial t} = -\vec{\nabla } \cdot \left(\sum_i \tau_{i,q} \vec{N}_\mathrm{i}^\mathrm{DM} \right) + {\dot{s}}_q\ \mathrm{.}
	\label{eq:continuity}
\end{equation}
Here, $\tau_{i,q}$ represents the stoichiometry of the solute $i$ in the quasi-particle $q$ and $\epsilon_\mathrm{e}$ the electrolytes volume fraction. 
The important feature of \cref{eq:continuity} is that the diffusion-migration flux of the quasi-particle is given by the weighted sum of the individual species $\vec{N}_\mathrm{q}^\mathrm{DM} = \sum_i \tau_{i,q}\vec{N}_\mathrm{i}^\mathrm{DM}$. 
The diffusion-migration flux of all individual species $\vec{N}_\mathrm{i}^\mathrm{DM}$ is calculated as
\begin{equation}
	\vec{N}_\mathrm{i}^\mathrm{DM} = \epsilon_\mathrm{e}^\beta D_i\vec{\nabla}c_i + \frac{t_i}{z_iF}\vec{j}\ \mathrm{,}
\end{equation}
where $D_i$ is the diffusion coefficient, $z_i$ the charge number, $t_i$ the transference number of the species, and $\vec{j}=-\kappa\vec\nabla\phi_\mathrm{elyt}$ is the current density as gradient of the electrolyte potential $\phi_\mathrm{elyt}$. 
We neglect the convection velocity \cite{Stamm2017, Kilchert2023}, as electrolyte volume in ZIBs remains approximately constant.
Electro-neutrality is enforced by the charge-conservation equation 
\begin{equation}
	0 = -\vec{\nabla}\cdot\vec{j} + \sum_i z_i \dot{s}_i^\mathrm{e}\ \mathrm{,}
\end{equation}
where $\dot{s}_i^\mathrm{e}$ is the source term due to the electrochemical reactions at the electrodes and is identical to the formulation in the regular Doyle-Fuller-Newman (DFN) models. 

\subsubsection*{Electrochemical and Precipitation Reactions}

Our continuum cell model contains the rates of the electrochemical half-cell reactions and the relevant precipitation reaction \cite{Blanc2020}.
These are the electrochemical dissolution and deposition of the metallic zinc anode, the electrochemical insertion reaction of both \ce{Zn^2+} and \ce{H^+}, the electrochemical dissolution of \ce{Zn_{0.5}MnO2}, and the precipitation of ZHS.

The zinc metal anode dissolves and reforms as redox reaction \cite{Schmitt2019},
\begin{equation}
	\ce{Zn <=> Zn^2+ + 2e-}\ \mathrm{.}
\end{equation}
This is the bare redox reaction. 
Upon solvation, the \ce{Zn^2+} ions form complexes in the electrolyte, e.g., \ce{ZnSO4}. 
Our quasi-particle formalism accounts for this formation of zinc-ligand complexes as discussed in \cref{ssec:qp}. 
We calculate the reaction rate of the \ce{Zn^2+} redox reaction using a symmetric Butler-Volmer rate,
\begin{equation}
	k_{\mathrm{ano}} = k^0_\mathrm{ano} \cdot \sqrt{\frac{c_{\ce{Zn^2+}}}{c_0}} \sinh{\left(\frac{zF}{2RT}\cdot\eta_\mathrm{ano}\right)}\ \mathrm{,}
\end{equation}
where $\eta_\mathrm{ano}$ is the overpotential at the anode surface, determined by the difference between electrode and electrolyte potential, i.e., $\eta_\mathrm{ano}=\phi_\mathrm{elde}-\phi_\mathrm{elyt} - \left(U_{0,\ce{Zn}} + \nicefrac{RT}{2F}\log\nicefrac{c_\mathrm{Zn^2+}}{c_0}\right)$.

The \ce{MnO2} cathode structure allows for the insertion of mono- and multivalent ions like \ce{H+} and \ce{Zn^2+}\cite{Juran2018}.
For the insertion of \ce{H+}, the reaction equation reads
\begin{equation}
	\ce{2 H+ + 2MnO2 + 2e- <=> 2 HMnO2}\ \mathrm{,}
	\label{eq:hinsertion}
\end{equation} 
and the insertion reaction of \ce{Zn^2+} is
\begin{equation}
	\ce{Zn^2+ + Mn2O4 + 2e- <=> ZnMn2O4}\ \mathrm{.}
\end{equation}
The corresponding Butler-Volmer rates are
\begin{equation}
	k_\mathrm{ins} = k^0_\mathrm{ins} \cdot 
	\sqrt{\mathrm{SOC}\cdot(1-\mathrm{SOC})\cdot\frac{c_i}{c_0}} \sinh{\left(\frac{z_iF}{2RT}
		\cdot\eta_\mathrm{ins}\right)}\ \mathrm{.}
	\label{eq:kins}
\end{equation}
Here, $c_i$ is the electrolyte concentration of the insertion species and
$\eta_\mathrm{ins}$ the corresponding overpotential. 
The exchange current density as prefactor depends on the state of charge ($\mathrm{SOC}$). 
We define it as $\mathrm{SOC} =  c_{i,\mathrm{solid}}/c_{i\mathrm{max}}$, where $c_{i,\mathrm{solid}}$ is the concentration of \ce{Zn} or \ce{H} in the cathode and $c_{i\mathrm{max}}$ is their maximal concentration in the material, i.e., \ce{HMnO2}, \ce{Zn_{0.5}MnO2}.
These Butler-Volmer equations are adapted from thermodynamical derivations
for the insertion reactions in Li-ion batteries \cite{Latz2013}.

Additionally, the electrochemical dissolution and deposition of \ce{Mn^2+} occur at the cathode. 
Based on DFT calculations presented in \cref{ssec:dft}, we will demonstrate that the dissolution of \ce{Zn_{0.5}MnO2} is the most relevant,
\begin{equation}
	\ce{2Zn_{0.5}MnO2 + 2e- + 8 H+ <=> Zn^2+ + 2Mn^2+ + 4H2O}\ \mathrm{.}
	\label{eq:mndiss}
\end{equation}
The open circuit voltage of this process $U_\mathrm{diss}$ is given as
\begin{align}
	U_\mathrm{diss} =& U_\mathrm{ref} \nonumber \\
	\quad& + \frac{RT}{zF}\left[ \log\left(
	\frac{c_{\ce{Zn^2+}}}{c_0}\cdot\frac{c_{\ce{Mn^{2+}}}}{c_0}^2\right) 
	-8\log(\frac{c_{\ce{H+}}}{c_0})\right]\ \mathrm{.}
	\label{eq:nernst_diss}
\end{align}
We model the dissolution and deposition rates for this reaction in analogy to the insertion reactions above,
\begin{align}
	k_\mathrm{diss} =& k^0_\mathrm{diss} \cdot \sqrt{\frac{c_{\ce{Zn_{0.5}MnO2}}}{c^\mathrm{max}_{\ce{Zn_{0.5}MnO2}}}} \nonumber \\
	\quad& \cdot \sinh\left(\frac{zF}{2RT}
	\cdot \left(\phi_\mathrm{cat}-\phi_\mathrm{elyt} 
	- U_\mathrm{diss}\right) \right)\ \mathrm{.}
	\label{eq:kdiss}
\end{align}
The equilibrium and kinetics of precipitation reactions depend on electrolyte pH. 
We will show in \cref{fig:ElectrolyteSpeciation} that ZHS is the only relevant precipitate in the cell studied here.
Thus, we include the precipitation of ZHS, a zinc-sulfate salt, in the cell model. 
The charge-neutral precipitation reaction of ZHS is given with
\begin{equation}
	\ce{4Zn^2+ + SO4^2- + 6 OH^- <=> Zn4(OH)6SO4 v}\ \mathrm{,}
	\label{eq:ZHS}
\end{equation}
which depends on pH through \ce{OH-} concentration.
Based on this, we calculate the saturation concentration for uncomplexed \ce{Zn^2+} with respect to ZHS precipitation as a function of pH and \ce{SO4^2-} as
\begin{equation}
	c_{\mathrm{sat}} = \left(K_{\mathrm{sp}} \cdot {c_{\ce{H+}}}^6 \cdot
	{c_{\ce{SO4^2-}}}^{-1}\right)^{\frac{1}{4}}\ \mathrm{,}
\end{equation}
with the solubility product $K_{\mathrm{sp}}$. 
We describe the dissolution reaction as a diffusion-limited process,
\begin{equation}
	k_\mathrm{prec}  = A_\mathrm{spec} D_{\ce{Zn^2+}} \epsilon^\beta 
	\cdot \frac{c_{\ce{Zn^2+}} - c_\mathrm{sat}}{\delta_0}\ \mathrm{,}
\end{equation}
with the diffusion layer thickness $\delta_0$. 
We model the nucleation process for ZHS with the oversaturation approach adopted from earlier works \cite{Horstmann2013} with the critical supersaturation ratio $s_\mathrm{critical}=\SI{105}{\percent}$ as for \ce{ZnO} in Ref. \citenum{Clark2017}.

\section{Simulation Results}
\label{sec:results}

In this section, we discuss the results of our calculations for the open circuit voltages of the electrodes, the equilibrium speciation and pH in the electrolyte, and the voltages during cycling.
First, we present the calculated energies of the \textdelta-\ce{MnO2*H2O} electrode structures and interpret their results for insertion and dissolution reaction (see \cref{ssec:dft}).
Following,  we use equilibrium thermodynamics to calculate the evolution of electrolyte pH and discuss the occurring precipitation reactions (see \cref{ssec:res_equil}).
We discuss the relevance of \ce{H+}-insertion into \ce{MnO2} during the first discharge phase.
In \cref{ssec:results_finalmodel},  we simulate the cell dynamics in both discharge phases, discuss the results of the transition to the second phase and the effects of ZHS precipitation at the cathode. 

\subsection{Electrode Potentials (DFT)}
\label{ssec:dft}

\begin{figure}[t]
	\begin{overpic}[width=\mycolumnwidth]{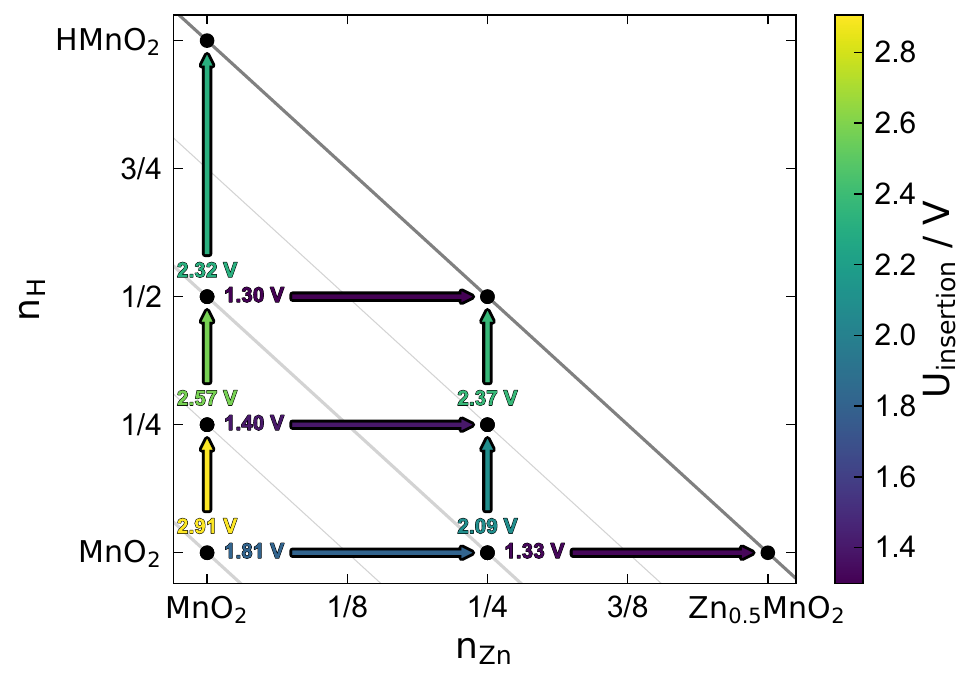}
		\put(40,45){\includegraphics[width=0.4\mycolumnwidth]{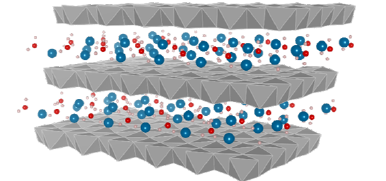}}
	\end{overpic}
	\caption{Insertion potentials calculated from DFT calculations.
		The electrochemical potentials for the dissolved species are evaluated for a $\SI{2}{\Molar}\,\ce{ZnSO4}$ + $\SI{.5}{\Molar}\,\ce{MnSO4}$ electrolyte.
		The inset in the upper right of the plot shows the crystal structure of \ce{Zn_{0.5}MnO2 * H2O}.}
	\label{fig:DFT_Potentials}
\end{figure}

\begin{table*}[ht]
	\center
	\caption{Electrochemical dissolution reactions and their potentials as given by the DFT calculations.}
	\begin{tabular}{@{}lll@{}}
		\toprule
		Dissolution reaction & $U_{\mathrm{diss}}$ @ ref. conditions & $U_{\mathrm{diss}}$ @ \ce{2M ZnSO4, 0.5M MnSO4}\\
		\midrule
		\ce{Zn_{$0.5$}MnO2 + 4H+ + 1e- <=> 0.5Zn^2+ + Mn^2+ + 2H2O}  & \SI{2.78}{\volt}  & \SI{1.71}{\volt}  \\
		\ce{Zn_{$0.25$}MnO2 + 4H+ + 1.5e- <=> 0.25Zn^2+ + Mn^2+ + 2H2O}  & \SI{2.29}{\volt}  & \SI{1.58}{\volt} \\
		\ce{MnO2 + 4H+ + 2e- <=> Mn^2+ + 2H2O}  & \SI{2.16}{\volt}  & \SI{1.63}{\volt}  \\
		\bottomrule
	\end{tabular}
	\label{tab:DFT_Diss_Potentials}
\end{table*}

The combination of chemical potentials of electrolyte species and the
structural energies of the \textdelta-\ce{MnO2} crystal structure allows
estimating the likelihood of relevant electrochemical structures and thereby reactions, namely \ce{Zn^2+}- or \ce{H+}-insertion and dissolution of the cathode structure. Therefore, we performed DFT calculations and thermodynamic calculations (see \cref{ssec:res_equil}) in order to calculate the corresponding open-circuit voltages (see \cref{ssec:theory_dft}).
Simulations of the structures for the proposed insertion states \cite{Zhao2020, Liao2022}, \ce{HMnO2 * H2O} and \ce{Zn_{0.5}Mn2O4 * H2O}, as well as a mixture of both, \ce{H_xZn_yMnO2 * H2O}, were executed and analyzed.
The calculated structures can be found in the Supporting Information.
By using \cref{eq:deltaG_DFT}, we calculate the theoretical insertion potentials for the distinct phases in a given environment.
We do this by using the chemical potentials of the dissolved species, calculated according to \cref{ssec:res_equil}.
In \cref{fig:DFT_Potentials}, the relative insertion potentials for the stepwise reactions at any stoichiometrically valid reaction are shown.
For pure \ce{H+}-insertion, we investigated the structures of \ce{H_{0.25}MnO2}, \ce{H_{0.5}MnO2} and \ce{HMnO2}.
Within a \SI{2}{\Molar} \ce{ZnSO4} + \SI{0.5}{\Molar} \ce{MnSO4} electrolyte, the insertion potential decreases from \SI{2.91}{\volt} to \SI{2.32}{\volt} at the end of the insertion process.
The investigated structures with solely \ce{Zn^2+}-insertion show insertion potentials between \SI{1.81}{\volt} to \SI{1.33}{\volt}. 
The insertion potentials for the \ce{H+}-insertion are greater at any point in the investigated phase space.

An electrochemical dissolution reaction of the \ce{MnO2}-cathode
is clearly observed in literature \cite{Qu1993,Wu2021, Guo2020} and on some occasions attributed as the key mechanism \cite{Chen2022, Chen2022a, Godeffroy2022} for the two phase behavior. 
We use the energies of formation acquired from DFT to evaluate the equilibrium potential for the dissolution reaction. 
The dissolution potential is given by the calculated $E_\mathrm{f}$ of the cathode, the chemical potential of the individual species in the electrolyte and the $E_\mathrm{f}$ for the bulk phase of \ce{Mn}, \ce{Zn} and \ce{H2(gas)}.
The results are listed in \cref{tab:DFT_Diss_Potentials}.
The energetically most favorable dissolution reaction is 2\ce{Zn_{0.5}MnO2 + 8H+ + 2e- <=> 2Zn^2+ + 2Mn^2+ + 4H2O} with a dissolution potential $U_\mathrm{diss} = \SI{1.71}{\volt}$. 
Its dissolution potential is larger than the average \ce{Zn^2+}-insertion potential.

In conclusion, the DFT calculations predict \ce{H+} insertion at the largest potentials, \ce{Zn_{0.5}MnO2} dissolution at intermediate potentials, and the \ce{Zn^2+} insertion at the lowest potentials.
The calculated potentials for \ce{Zn^2+} insertion and \ce{Zn_{0.5}MnO2} dissolution agree nicely to the potential range observed in experiments \cite{Alfaruqi2015c,Ren2019,Chen2022,Chen2022a}.
However, the potential for \ce{H+} insertion seems very large compared to observed cell voltages and the stability window of aqueous electrolytes.

\subsection{Electrolyte Speciation (Thermodynamics)}
\label{ssec:res_equil}

\begin{figure}
	\includegraphics[width=\mycolumnwidth]{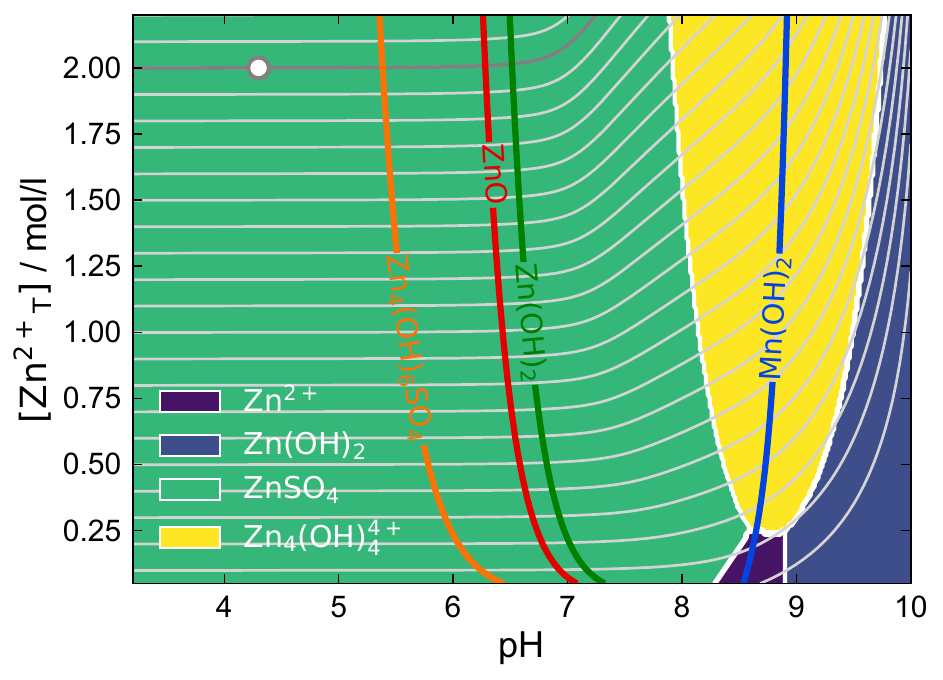}
	\caption{Phase diagram of electrolyte speciation and precipitation reaction for the \ce{ZnSO4} electrolyte with \SI{.5}{\Molar} \ce{MnSO4} additive. 
		The background colors depict the dominant aqueous zinc complexes. 
		The solid-colored lines correspond to the solubility of the respective precipitates.
		The solid gray lines show paths of constant $[\ce{SO4^2-}_\mathrm{T}]$, which are invariant with respect to electrochemical reactions. The white circle indicates the initial state of a benign solution and the dark gray line its corresponding isoline.}
	\label{fig:ElectrolyteSpeciation}
\end{figure}

Calculating the equilibrium speciation in the \ce{ZnSO4} electrolyte gives us an overview of the dynamics of electrolyte pH and precipitation products.
The law of mass action presented in \cref{ssec:qp} determines the speciation in the aqueous electrolyte.
\Cref{fig:ElectrolyteSpeciation} shows the dominant zinc-ligand complex as a function of pH and total zinc concentration for the fixed manganese and sulfate concentrations $[\ce{SO4^2-}_\mathrm{T}] = \SI{2.5}{\Molar}$ and $[\ce{Mn^2+}_\mathrm{T}] = \SI{.5}{\Molar}$.
We demonstrate the distribution concentrations of all \ce{Zn^2+}-complexes in the Supporting Information in \cref{SI-fig:ElectrolyteSpeciation}. 
The solid gray lines are paths of constant $[\ce{SO4^2-}_\mathrm{T}]$-concentration, which is invariant under the electrochemical reactions.
Precipitation of ZHS, \ce{ZnO}, \ce{Zn(OH)2}, and \ce{Mn(OH)2} occurs in the region above the colored lines.

If a ZIB battery is discharged slowly, the electrolyte will stay homogeneous throughout the cell.
The initial pH of our electrolyte is $\approx \num{4.3}$, the total zinc concentration is $[\ce{Zn}_\mathrm{T}]=\SI{2}{\Molar}$, indicated by the white dot in \Cref{fig:ElectrolyteSpeciation}.
During discharge, electrolyte pH is expected to rise due to \ce{H+} insertion (see \cref{eq:hinsertion}) and \ce{Zn_{0.5}MnO2} dissolution (see \cref{eq:mndiss}).
Thus, the electrolyte state during discharge follows the solid gray lines in \cref{fig:ElectrolyteSpeciation} towards higher pH.
When the electrolyte state hits the orange line, ZHS precipitation starts.
The precipitation of ZHS reduces the amounts of $[\ce{SO4^2-}_\mathrm{T}]$ and $[\ce{OH^-}]$ (see \cref{eq:ZHS}) buffering the pH. 

To conclude, the relevant pH region for the studied ZIB is below \num{6} and the dominant species during operation is in all cases a neutral \ce{Zn^2+}-\ce{SO4^2-} complex.
We conclude that the precipitation of ZHS is thermodynamically favorable.
The precipitation of zinc or manganese (hydro)oxides is not expected.
The solubility limit of ZHS is reached at a pH of \numrange{5.2}{6.3} depending on the \ce{Zn^2+}-concentration.

The maximal pH change before the onset of ZHS precipitation must correlate with the capacity for \ce{H+} insertion (see \cref{eq:hinsertion}) and \ce{MnO2} dissolution (see \cref{eq:mndiss}). 
As \ce{H+} insertion is often assumed to occur at the beginning of discharge and we calculate a large OCV in \cref{ssec:dft}, we focus on \ce{H+} insertion here and defer the analogous discussion for \ce{MnO2} dissolution to the SI (see \cref{SI-fig:ElectrolyteStabilityMn}).
Our goal is to identify reaction mechanisms by quantifying this effect and comparing it with experimentally observed capacities.

\begin{figure}[b]
	\includegraphics[width=\mycolumnwidth]{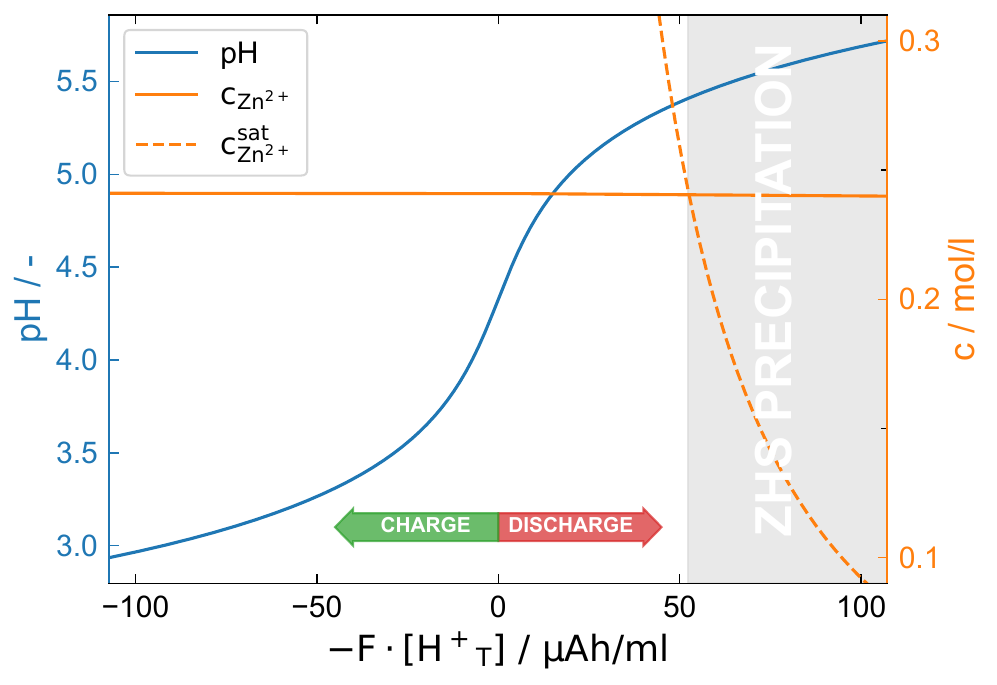}\hfill
	\caption{Dependence of electrolyte pH and \ce{Zn^2+} saturation with respect to ZHS precipitation as a function of proton concentration for a \ce{H^+} reactions in a \SI{2}{\Molar} \ce{ZnSO4}, \SI{0.5}{\Molar} \ce{MnSO4} electrolyte.
		The pH is shown on the left, and zinc concentration and zinc saturation concentration are shown on the right; both are shown as a function of the amount of \ce{H+} added to the electrolyte. 
		We argue that the insertion of \ce{H+} into \ce{MnO2} during discharge would result in an identical decrease of $[\ce{H+}_\mathrm{T}]$ in the electrolyte.}
	\label{fig:ElectrolyteStability}
\end{figure}

We simulate how the electrolyte composition and pH value are influenced by a \ce{H+}-insertion at the cathode, which reduces the $[\ce{H^+}_T]$ concentration in the electrolyte (see \cref{eq:hinsertion}).
As the \ce{H+} insertion is balanced by the zinc metal dissolution, $-2\cdot\Delta[\ce{H+}_\mathrm{T}] = \Delta[\ce{Zn^2+}_\mathrm{T}]$ holds.
\cref{fig:ElectrolyteStability} shows the pH value and \ce{Zn^2+} saturation limit in dependence on the total amount of $\ce{H+}$ in the electrolyte. We show the results of an equivalent simulation for \ce{Mn^2+} dissolution in the Supporting Information (\cref{SI-fig:ElectrolyteStabilityMn}).

We find that the pH around equilibrium is highly sensitive to the \ce{H+} insertion reaction.
The saturation limit is reached after \SI{2}{\micro\mol\per\liter} of \ce{H+} are inserted into the electrolyte which equals a discharged capacity of \SI{52}{\micro\ampere\hour\per\milli\liter}.
In typical laboratory coin cells, reported electrolyte to active mass ratios are in the order of \SI{30}{\milli\liter\per\gram} \cite{Ren2019}. When we now use our calculations to estimate ZHS onset for the experimental electrolyte to active mass ratios, the onset is expected after approximately \SI{0.15}{\milli\ampere\hour\per\gram}. 
However, the precipitation of ZHS is experimentally observed after a discharged capacity greater than $\SI{100}{\milli\ampere\hour\per\gram}$.

\begin{figure*}[!t]
	\subfigure{
		\begin{overpic}[width=0.4697\textwidth]{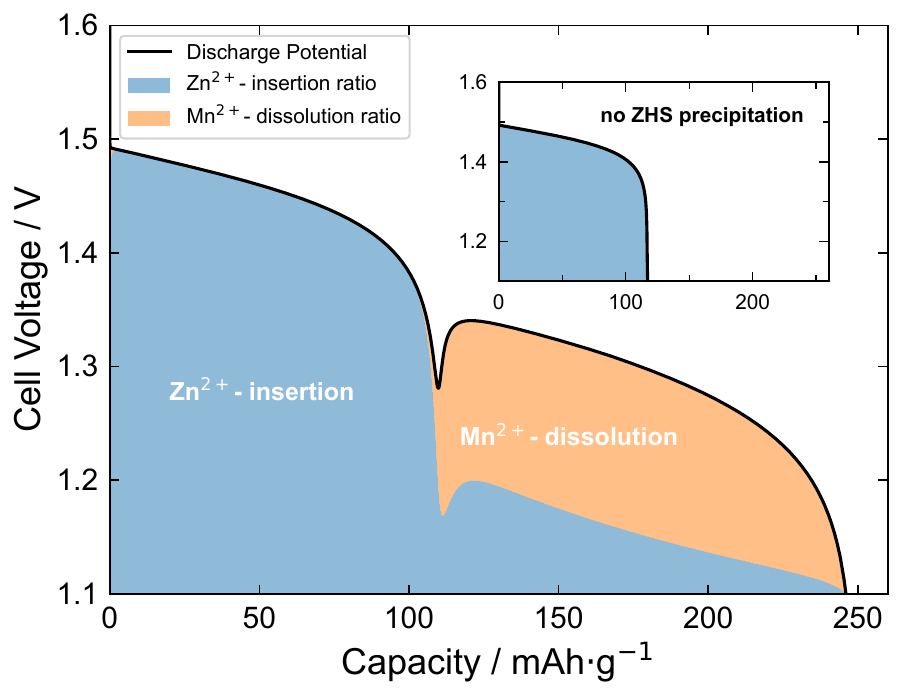}
			\put(1,70){a)}
		\end{overpic}
	}\hfill
	\subfigure{
		\begin{overpic}[,width=0.5103\textwidth]{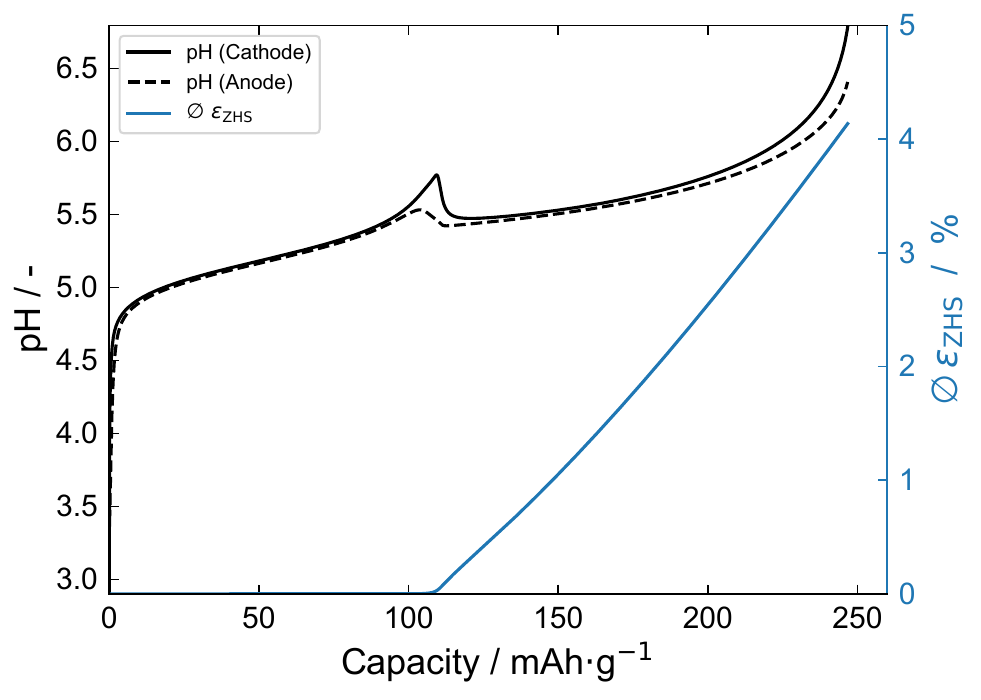}
			\put(1,64){b)}
		\end{overpic}
	}
	\caption{a) Galvanostatic discharge behavior at \SI{200}{\milli\ampere\per\gram} (equal to \SI{0.4}{\milli\ampere\per\cm\squared} at the simulated mass loading of \SI{2}{\milli\gram\per\cm\squared}). Shown are the discharge voltages based on \ce{Zn^2+}-insertion and \ce{Mn^2+} dissolution. The main axis shows simulations including the ZHS precipitation reaction, while the simulation shown in the inset neglects this reaction. Only the full model reproduces the second discharge phase.
		The colored areas below the discharge potential represent the fractional contribution of the \ce{Zn^2+}-insertion and \ce{Mn^2+}-dissolution to the cell current. Here, \ce{Mn^2+}-dissolution becomes significant only in the second discharge phase. b) Dynamics of electrolyte pH and ZHS precipitation for the full cell model. Shown are the electrolyte pH at both anode and cathode as well as the average volume-fraction of ZHS in the cell. While there is 
		a pH increase in both discharge phases, ZHS growth happens only in the second discharge phase. The pH at the end of the first discharge phase sharply increases but is lowered again once ZHS growth starts.}
	\label{fig:mndiss_fundamental_sim}
\end{figure*}

Thus, the first discharge phase cannot be dominated by \ce{H+}-insertion (or \ce{MnO2} dissolution) as experiments find no ZHS precipitation in this phase. 
In the non-equilibrium case for discharge at realistic rates, diffusion limitations further accelerate local pH change and ZHS precipitation as we discuss in detail in the Supporting Information in \cref{SI-ssec:results_firstphase}.
In combination with the calculated electrode potentials (see \cref{ssec:dft}), we conclude that the \ce{H+}-insertion reaction, even if it is energetically favorable, must be strongly kinetically suppressed and is not relevant for the cycling mechanism found in \ce{MnO2}-based ZIBs.

\subsection{Discharge Phases (Cell Model)}
\label{ssec:results_finalmodel}

As discussed above, the insertion of \ce{H+} into \ce{MnO2} cathode cannot dominate the first discharge phase.
Thus, we model the discharge with the combination of \ce{Zn^2+}-insertion and \ce{Mn^2+} dissolution at the cathode.
We simulate the galvanostatic discharge of a laboratory coin cell in the presence of ZHS precipitation and plot the discharge voltage in \cref{fig:mndiss_fundamental_sim}a.
The cell voltage shows two discharge phases with a voltage dip in between as generally reported in the literature \cite{Borchers2021}.
The filled regions below the discharge curve represent the relative contribution of the \ce{Zn^2+}-insertion and \ce{Mn^2+} dissolution reaction.
The first discharge phase is dominated by the insertion of \ce{Zn^2+}.
The \ce{Mn^2+} dissolution onsets shortly before the voltage dip and becomes relevant in the second discharge region. 
This is in excellent agreement with the experimental findings of Wu et al. \cite{Wu2018} that the \ce{Mn^2+} content in the electrolyte significantly increases in the second phase.

In the inset of \cref{fig:mndiss_fundamental_sim}a, we neglect ZHS precipitation for comparison.
In this case, only the first discharge phase is present and the contribution of \ce{Mn^2+} dissolution is negligible.
Thus, ZHS precipitation is required to reproduce the two distinct discharge phases.

\begin{figure}[!b]
	\includegraphics[width=\mycolumnwidth]{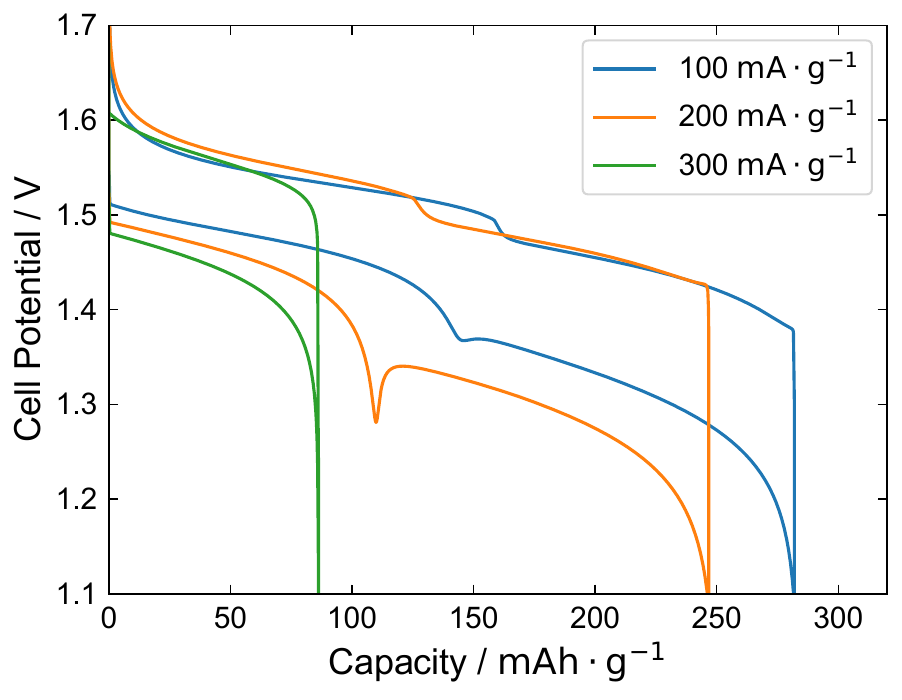}
	\caption{Cycling voltage for current rates of \SI{100}{\milli\ampere\per\gram}, \SI{200}{\milli\ampere\per\gram}, and \SI{300}{\milli\ampere\per\gram}. Shown are the cell potentials during galvanostatic discharge and charge during the second cycle. At low current densities, the second discharge phase is clearly defined and the phase distinction is also visible during charging. At the highest rate, the voltage dip and the additional capacity of the second phase are not present.}
	\label{fig:mndiss_rates_sim}
\end{figure}

\begin{figure*}
	\includegraphics[width=\textwidth]{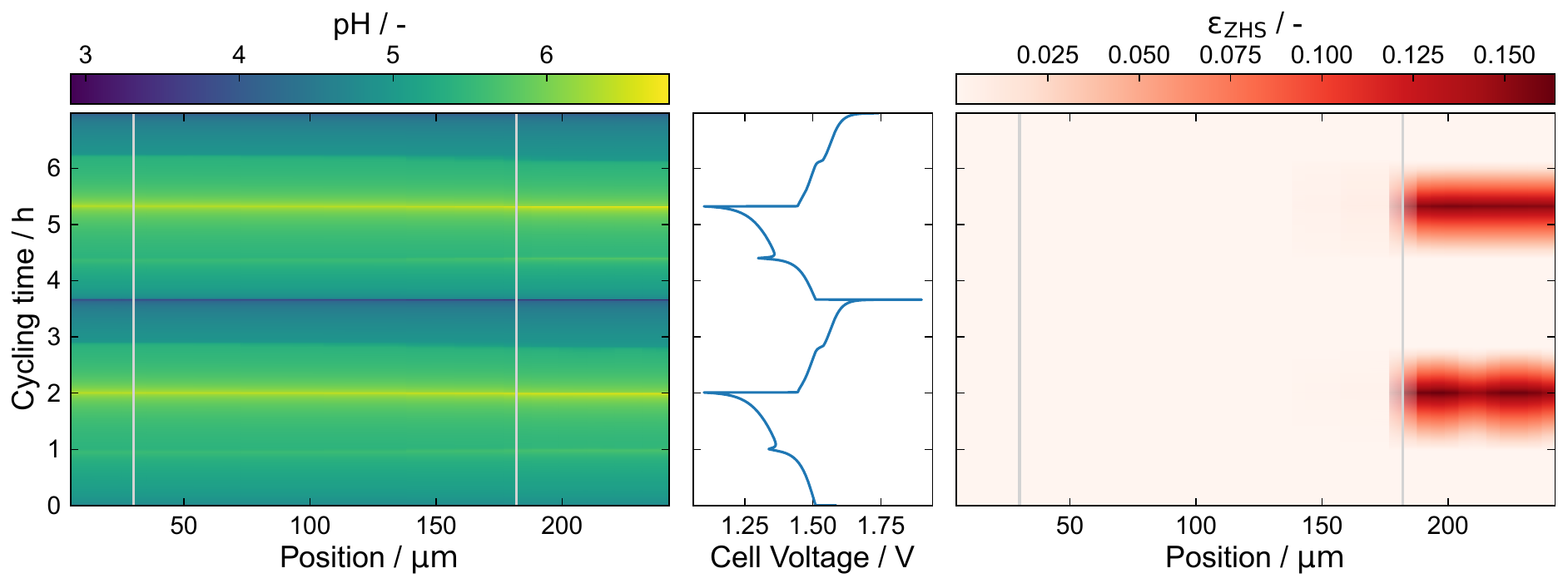}
	\caption{Electrolyte pH and volume fraction of the precipitate $\mathrm{\epsilon_{ZHS}}$ over two cycles.
		During discharge, electrolyte pH increases gradually from an initial value of $\approx\num{4.3}$.
		Around the dip in voltage, the pH reaches values that are higher than the saturation limits
		but drops with the precipitation of ZHS. In the second half of the discharge, the
		volume of $\epsilon_\mathrm{ZHS}$ grows while the pH remains mostly constant throughout the cell.
		At the end of the charge, the sharper voltage decline is associated with a more rapid increase of the pH.}
	\label{fig:spatial_eps_pH}
\end{figure*}

Thus, we analyze electrolyte pH and average ZHS volume fraction $\epsilon_\mathrm{ZHS}$ during discharge in \cref{fig:mndiss_fundamental_sim}b.
ZHS precipitation is limited to the second discharge phase.
Its onset is correlated with the voltage dip in \cref{fig:mndiss_fundamental_sim}a.
The pH value increases during the first phase of discharge. 
At the end of the first phase, the pH rises sharply until it is reduced again at the onset of ZHS precipitation. 
During the second phase, electrolyte pH slowly increases at first before strongly raising near the end of discharge.

We can rationalize this behavior based on the chemical reactions for \ce{Zn_{0.5}MnO2} dissolution and \ce{Zn^2+} insertion.
During discharge, \ce{Zn_{0.5}MnO2} dissolution releases \ce{OH-} into the electrolyte (see \cref{eq:mndiss}) so that the electrolyte becomes more alkaline.
While the rate of \ce{Zn^2+} insertion is independent of pH, the equilibrium voltage of \ce{Zn_{0.5}MnO2} dissolution (see \cref{eq:nernst_diss}) strongly decreases with electrolyte pH $U_\mathrm{diss}\approx U_\mathrm{diss}^0 +  238\mathrm{mV}\cdot\mathrm{pH}$.
Consequently, the dissolution potential for this reaction drops.
In turn, the pH increase limits the dissolution reaction as long as the cell voltage is stabilized by \ce{Zn^2+} insertion.
When \ce{Zn^2+} insertion becomes more difficult due to transport limitations in the \ce{MnO2} material, the cell voltage drops and \ce{Zn_{0.5}MnO2} dissolution accelerates.
As a consequence, the pH value increases quickly and ZHS starts to precipitate (see \cref{fig:ElectrolyteStability}).
The pH-driven precipitation (see \cref{fig:ElectrolyteSpeciation}) removes \ce{OH-} from the electrolyte and stabilizes the pH near its saturation limit.
ZHS precipitation makes possible significant \ce{Zn_{0.5}MnO2} dissolution without its self-limiting mechanism.

This interplay between ZHS precipitation and \ce{Zn_{0.5}MnO2} dissolution, which is first described by our theory, is key to our consistent model of the cycling mechanism of \ce{MnO2}-based ZIBs.
The \ce{Zn_{0.5}MnO2} dissolution, while energetically more favorable, is a self-limiting reaction in the first discharge phase without precipitation.
The onset of ZHS precipitation, observed as a nucleation dip in cell voltage, stabilizes electrolyte pH and resolves the self-limitation of pure \ce{Zn_{0.5}MnO2} dissolution.
In the second discharge phase, the dissolution of the \ce{Zn^2+}-inserted manganese oxide \ce{Zn_{0.5}MnO2} contributes significantly to the overall capacity and drives the precipitation of ZHS.
During charging, the ZHS will be dissolved again and the cathode is redeposited.
Laboratory ZIBs are often optimized with respect to capacity and thus designed towards significant \ce{Zn_{0.5}MnO2} dissolution. 
However conversion electrodes are prone to shape change \cite{Schmitt2019} and the deposition process of \ce{MnO2}-structures can change its crystal structure \cite{Alfaruqi2018a}.
Thus, we expect that this common optimization strategy limits cycle life and induces accelerated aging.
We propose to reduce \ce{Zn_{0.5}MnO2} dissolution to reduce aging and capacity fade. 
In \cref{ssec:optimization}, we optimize the discharging strategy towards this rationale.

To gain further insights, we simulate the cycling behavior for different current densities.
We simulate several cycles with a galvanostatic charge and discharge, both at the same current density.
\cref{fig:mndiss_rates_sim}, shows the charge and discharge potentials of the second cycle.
During charging, we find two clearly separated phases without a separating voltage dip.
The voltage dip between the phases is present at low currents but disappears at higher rates.
The contribution of the second phase is decreasing with increasing currents and is fully suppressed at high currents.
This shows how sensitive the voltage reacts to variations in cycling currents.
In turn, small differences in material preparation and cell design can also strongly affect cell behavior.

\section{Discussion}
\label{sec:discussion}

In the following section, we compare the behavior of our theory-based model with experimental observations from the literature to validate our approach.
Hereby, we compare the (dis)charge voltages, investigate the tempo-spatial profiles of pH evolution as well as precipitation within our cell model and present the results of cyclovoltammetry simulations in \cref{ssec:validation}.
Subsequently, we discuss strategies to increase cycling stability and reduce \ce{MnO2} dissolution and ZHS precipitation by adding \ce{MnSO4} into the electrolyte, by increasing electrolyte volume, and by adjusting the cycling protocol (see \cref{ssec:optimization}). 

\subsection{Validation}
\label{ssec:validation}

We use literature data of measured cell potentials during cycling to validate our proposed cycling mechanism. 
Experimental results show two phases during discharge, separated by a voltage dip, which is reproduced by our model.
A comparison of experimental discharge voltages of \textdelta-\ce{MnO2} coin cells, as found in References \citenum{Alfaruqi2015c, Chen2022, Chen2022a, Ren2019}, with our simulation results is plotted in the Supporting Information in \cref{SI-fig:LitDischarge}.
The experiments show the same discharge and charge behavior as our simulations, with two phases that are separated by a voltage dip during discharge (see \cref{fig:mndiss_rates_sim}).
Observed rate dependencies of the cycling behavior for \textdelta-\ce{MnO2} as, for example, investigated by Guo et. al. \cite{Guo2019} and Ren et. al. \cite{Ren2019} show that the second-phase capacity is reduced significantly with increasing current densities.
At high rates, it is also observed that the second phase might not even occur. 
This behaviour is similarly observed for other \ce{MnO2} polymorphs, e.g., for \textalpha-\ce{MnO2} \cite{Wu2018}, \textepsilon-\ce{MnO2} \cite{Huang2021a}, and amorphous \ce{MnO2} \cite{Bi2020}.
In our computational study, we find that the second phase disappears at higher rates due to the slow kinetics of the precipitation reaction (see \cref{fig:mndiss_rates_sim}).
Quantitative differences between our simulation and the different lab cell measurements are a result of different synthesis approaches, cell design and applied current. 
We summarize that our model reproduces the key experimental features, i.e., the two discharge phases, the voltage dip, as well as the rate dependence of the two phases.

The evolution of ZHS is measured by Putro et. al \cite{Putro2020} and Chen et. al. \cite{Chen2022}.  
Their in-situ spectroscopy data show a reversible growth and dissolution of ZHS during cycling, which is occurring in the second phase of discharge and the first phase of charge \cite{Putro2020,Chen2022}. 
The right subfigure of \cref{fig:spatial_eps_pH} presents our simulation results for ZHS volume for two consecutive cycles of the cell model fraction in a spatially resolved way.
Our simulations nicely reproduce these experimental findings for ZHS growth.
We also find that ZHS precipitation occurs in the cathode only and does not extend into the anode.

In 2016, Lee et. al. \cite{Lee2016a} investigated the pH evolution in a ZIB with a $\alpha$-\ce{MnO2} cathode during the first cycle. 
We compare our simulations with recent investigations of Biro and coworkers \cite{Bischoff2020, Fitz2021}. 
They study in detail the pH evolution over several cycles and find that the pH evolution is reversible.  
Electrolyte pH is measured separately in the anode and cathode.
They highlight a sharp decrease in pH at the end of the charge.
The left subfigure of \cref{fig:spatial_eps_pH} shows our simulation results for pH evolution within the active region of the cell. 
Our model reproduces the reversible behavior of electrolyte pH and the sharp increase at the end of discharge found by Biro and coworkers \cite{Bischoff2020, Fitz2021}.
Our simulations predict no significant pH gradient between the cathode and anode because our coin-cell geometry is significantly smaller than the laboratory setup of Biro and coworkers granting space for the pH measurement device.  
In combination with the excellent conductivity of aqueous electrolytes, the rather uniform pH distribution is according to our expectations.  

\begin{figure}[t]
	\includegraphics[width=\mycolumnwidth]{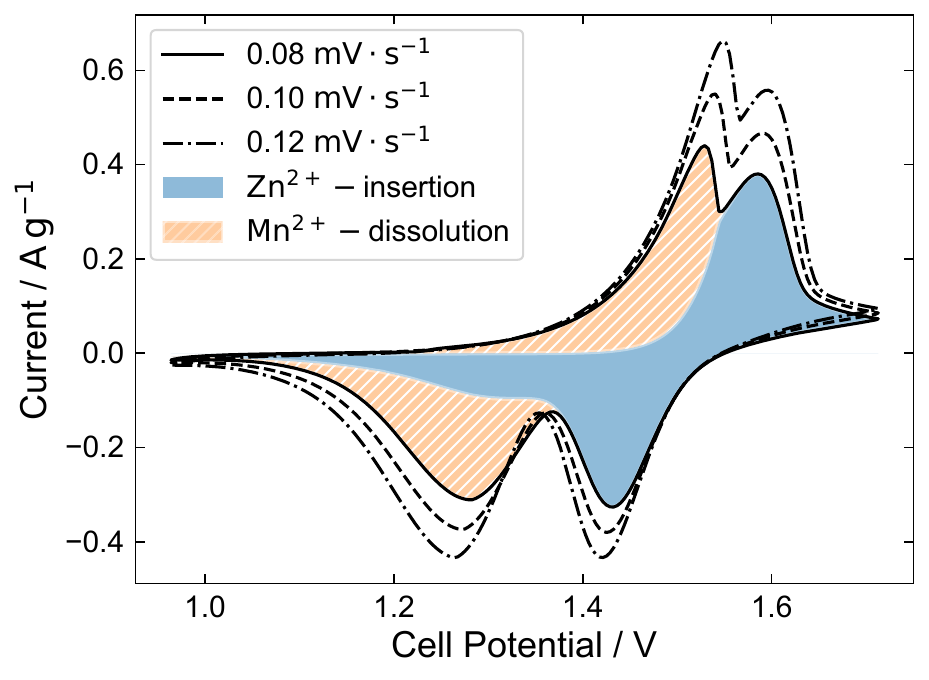}
	\caption{Simulated voltammetry measurements. The current density is shown as a function of applied potential for sweep rates from
		\SIrange{0.08}{0.12}{\milli\volt\per\second} as black lines. The colored 
		regions below the current curve show the current contributions of the 
		\ce{Zn^2+}-insertion and \ce{Mn^2+}-dissolution reaction.
		The first discharge-peak is dominated by the \ce{Zn^2+}-insertion reaction,
		\ce{Mn^2+}-dissolution is only relevant in the second peak.} 
	\label{fig:cv}
\end{figure}

Cyclovoltammograms (CVs) are used in experiments to identify individual processes by their characteristic redox peaks.
We perform cell simulations and elucidate the direct correlation of the characteristic of the CVs for \ce{MnO2}-cathodes with the underlying electrochemical reaction.
\cref{fig:cv} shows the simulated cyclovoltammograms of our cell model.
We observe that the experimentally described separation of two redox peaks \cite{Qu1993, Ren2019, Guo2019} is predicted by our cell model. 
The filled areas in \cref{fig:cv} visualize how the rates of the individual electrochemical reactions at the cathode contribute to the overall cell current.
Here, the first peak in discharge directions can be associated with the \ce{Zn^2+} insertion reaction, while the second discharge peak is a result of the onset of the \ce{Mn^2+} dissolution.
In the charging direction, the \ce{Mn^2+} of the cathode is redeposited first as \ce{Zn_{0.5}MnO2} and then the remaining \ce{Zn^2+} is de-inserted. 

\begin{figure}[t]
	\includegraphics[width=\mycolumnwidth]{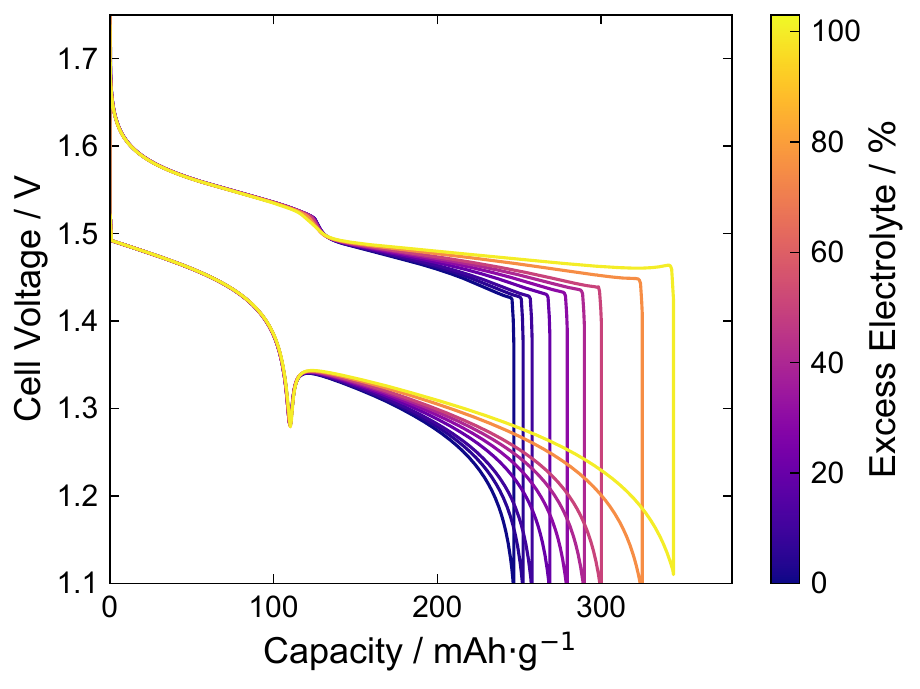}
	\caption{Discharge behavior with different electrolyte volumes. 
		The amount of electrolyte is increased relative to the minimal volume used 
		to wet electrodes and separator by up to \SI{30}{\percent}. The end of the first discharge phase is hardly influenced by excess electrolyte (compare inset), while the 
		second discharge phase becomes longer, the more electrolyte is added to the cell.}
	\label{fig:elytvolume}
\end{figure}

\subsection{Optimization}
\label{ssec:optimization}
Based on our cell model we investigate strategies to reduce \ce{MnO2} dissolution and ZHS precipitation.
In this section, we discuss the effect of \ce{MnSO4} as electrolyte additive and electrolyte volume variations.
Finally, we present a modified discharge protocol that allows for improving the insertion/dissolution ratio.

The volume of the electrolyte influences pH stability and changes the precipitation dynamics of ZHS.
Our calculations in \cref{ssec:res_equil} showcase the sensitivity to excess electrolyte on a pH-driven precipitation reaction.
In \cref{fig:elytvolume}, we present a study of cycling behavior for different electrolyte volumes based on our \ce{Zn^2+}-insertion/\ce{Mn^2+}-dissolution model.
We implement a reservoir with excess electrolyte and increase the electrolyte amount, starting from \SI{9.2}{\micro\liter\per\cm\squared}, which is the amount needed to fill the pore volume in the anode, separator and cathode. 
While the capacity of the first discharge phase, which is dominated by the \ce{Zn^2+} insertion process, is rarely influenced by the amount of excess electrolyte in \cref{fig:elytvolume}, the \ce{Mn^2+}-dissolution phase is significantly extended in the presence of more electrolyte.
We conclude that the ZHS precipitation/\ce{Mn^2+} dissolution mechanism is sensitive to ion depletion in small electrolyte volumes.  

\begin{figure}[t]
	\includegraphics[width=0.884\mycolumnwidth]{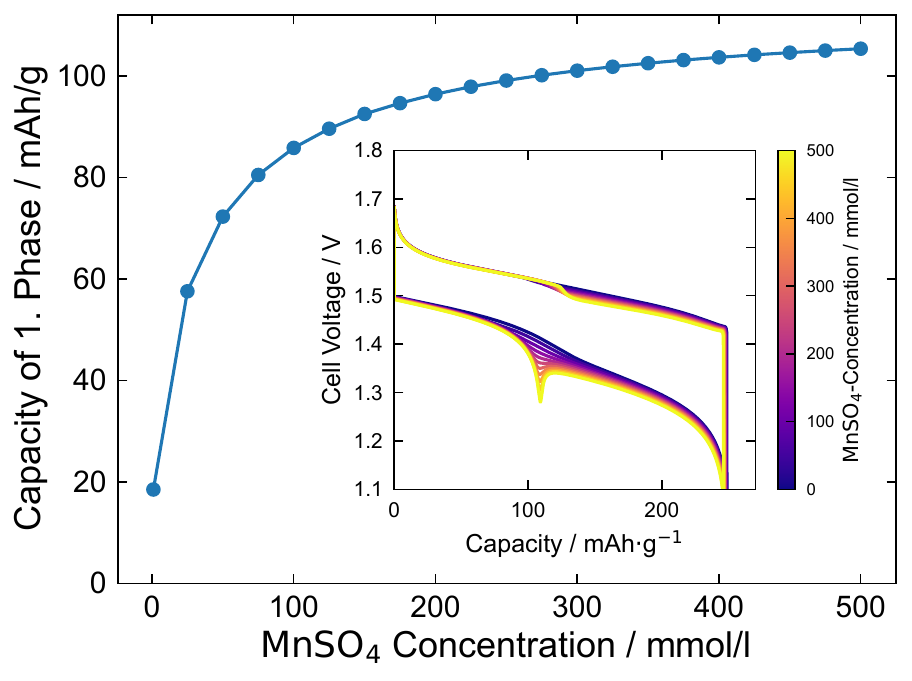}
	\caption{Discharge behaviour with different amounts of \ce{MnSO4} pre-added to the electrolyte. Shown is the quantitative analysis of the \ce{Mn^2+}-additive. The major axis displays the first phase capacity as a function of pre-added \ce{MnSO4}. The inset axis shows the charge and discharge behavior in the second cycle. The higher the amount of \ce{MnSO4}-additive, the sharper the transition between the first and second discharge phases. While the onset of the second phase is significantly deferred with \ce{MnSO4}-additive, the capacity is hardly influenced.}
	\label{fig:mnaddition}
\end{figure}

\ce{MnSO4} is often used as electrolyte additive in order to inhibit \ce{MnO2} dissolution \cite{Kim1998,Chamoun2018a,Borchers2021}.
The amount of pre-added \ce{MnSO4} is mostly empirically motivated.
While early work of Kim et. al. \cite{Kim1998} showed optimum cycling stability for \SI{0.1}{\Molar} \ce{MnSO4}, the recent work of Chen et. al. \cite{Chen2022} uses \SI{0.5}{\Molar} \ce{MnSO4}.
\cref{fig:mnaddition} presents a comparison of \ce{MnSO4}-influence on cycling performance.
In the inset of \cref{fig:mnaddition}, the cell voltage during cycling is shown.
While the achievable capacity is only slightly dependent on the \ce{MnSO4} amount, larger amounts of \ce{MnSO4} result in more pronounced voltage dips associated with the nucleation of ZHS.
The main part of \cref{fig:mnaddition} evaluates the capacity at which ZHS precipitation is first observed.
We find that the onset of the second phase with \ce{MnO2} dissolution occurs later if larger amounts of \ce{MnSO4} are pre-added. 
In summary, \ce{MnSO4}-additive effectively allows for a significantly larger discharge capacity in the first phase.
Evaluation of the ratio of capacity from the \ce{Zn^2+}-insertion and capacity from \ce{Mn^2+}-dissolution gives a \ce{Zn^2+} contribution of $\approx \SI{62}{\percent}$ for a discharge at \SI{2}{\ampere\per\meter\squared}, which is in agreement with the experimental findings of Yang et. al. \cite{Chen2022a}. 
However, the change of this ratio is less than \SI{1}{\percent} for cycling in pure \ce{ZnSO4} as compared to the electrolyte with \SI{0.5}{\Molar} \ce{MnSO4}.
We therefore find that \ce{MnSO4} helps to prolong the first phase, but does not significantly change the total discharge capacity and the relative contribution of the \ce{MnSO4}-dissolution process.

\begin{figure}[t]
	\includegraphics[width=\mycolumnwidth]{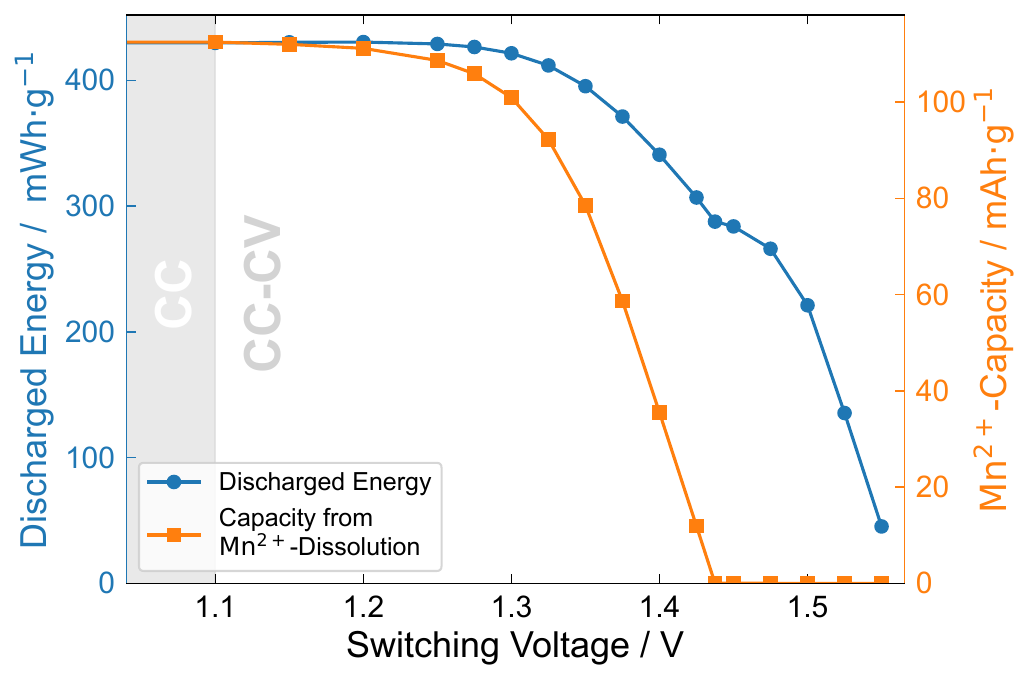}
	\caption{Optimized Discharge Performance with a CC-CV-type discharge. The contribution of capacity from the \ce{Mn^2+}-process and the achievable overall energy are presented in relation to their values at a standard CC discharge. The values are shown as a function of the switching voltage between CC and CV discharge. If the switching voltage 
		is higher than the cell voltage at the start of the second phase, the dissolution process is significantly suppressed.}
	\label{fig:op}
\end{figure}

Recently published works on high-performance ZIBs all salvage the additional capacity achievable in the second discharge phase which is associated with cathodic dissolution\cite{Yang2023}. 
However, experimental studies also report crystallographic changes in redeposited \ce{MnO2} during charging \cite{Lee2014, Huang2019, Wang2019}.
Additionally, dissolution and redeposition of the \ce{MnO2}-structure has been claimed to be a reason for reduced cycle life \cite{Liang2021}.
Therefore, limiting the cathode dissolution might help achieve higher cycling stability.
\cref{fig:op} shows the influence of a constant current-constant voltage (CC-CV) discharge profile on the achievable energy.
We conducted discharge simulations with a constant current at the start, once a certain voltage is reached, the discharge is switched to potentiostatic mode.
We varied the switching voltage between \SIrange{1.1}{1.55}{\volt}.
Here, we find that switching discharge modes from galvanostatic to potentiostatic mode has a significant leveraging effect on the cathodic dissolution.
At the switching region around \SIrange{1.3}{1.4}{\volt}, the cathode dissolution can be suppressed without sacrificing any of the capacity of the \ce{Zn^2+}-insertion process.

\section{Conclusion}

This article discusses the relevance of proposed reaction mechanisms in the \ce{MnO2} cathode in \ce{ZnSO4} electrolyte, i.e., \ce{H+} insertion, \ce{Zn2+} insertion, and \ce{MnO2} dissolution.
The calculated electrode potentials by DFT indicate that a \ce{H+} insertion reaction is energetically more favorable.
Based on calculations for electrolyte thermodynamics, however, we conclude that a \ce{H+} consuming reaction can not be dominant in the first half of discharge.
Contrary to the expectations from \ce{MnO2}-cathodes in alkaline electrolytes, the first discharge phase is thus dominated by the insertion of \ce{Zn^2+} ions. 

The continuum cell model for ZIB cells with \ce{MnO2} cathodes developed in this work reproduces the two-phase cycling behavior.
It is used to investigate the critical role of ZHS precipitation for the second discharge phase.
This work proposes feedback between the cathode's electrochemical dissolution and the stabilizing effect of ZHS dissolution on electrolyte pH.
With the nucleation of ZHS, electrolyte pH is stabilized at the saturation limits which allows for continuous \ce{MnO2} dissolution.
Validated by different in-situ experiments, our simulation results show that the developed theory with its pH-based feedback process can reproduce the two-phase cycling characteristics of \ce{MnO2}-based ZIBs and the double-peak structure in cyclovoltammetry measurements.
The unique voltage dip during discharging is identified as a result of the nucleation of ZHS at the cathode.

With this consistent understanding of the cycling mechanism, theory-based optimization strategies become possible.
The combination of conversion reactions, i.e., \ce{MnO2} dissolution and ZHS precipitation, increases discharge capacity, but leads to shape change and capacity fade during continued cycling.
We present an optimized CC-CV-discharging protocol, which can mitigate cathode dissolution also at low current densities.
Another optimization approach would be electrolyte design based on our theoretical expectations, such as suppression of the ZHS stabilizing mechanism.

\section{Computational Section}
\label{sec:compsec}

Periodic density functional theory (DFT) calculations were performed to investigate the proton and zinc insertion in \textdelta-\ce{MnO2}.
For this purpose, the Vienna ab initio simulation package (VASP) was applied, using the Projector Augmented Wave (PAW) method to describe the electron-core interaction \cite{Kresse1993,Kresse1996,Kresse1999, Blochl1994a}.
While exchange and correlation were accounted for by the generalized gradient approximation in the formulation of Perdew, Burke and Ernzernhof (PBE) an additional Hubbard-like correction - with a U parameter of \num{3.9} - was included to describe the localized character of the Mn d-electrons \cite{Perdew1996, Jain2011}.
All calculations were based on supercells of a 9 atom \textdelta-\ce{MnO2} cell that contained one water molecule, i.e., \ce{Mn2O4 * H2O}, using an energy cutoff of \SI{600}{\electronvolt} and a 7x14x5 K-point mesh for the unit cell, which was adapted accordingly for larger supercells.
To investigate possible intercalation compounds, different numbers of Zn and H atoms were inserted in the respective supercells, corresponding to \ce{H_xZn_yMnO2*H2O} stoichiometries (with x and y equal to 0, 0.25, 0.5 and 1).
The structures were relaxed with respect to lattice vectors and atomic positions, applying convergence criteria of \SI{10e-6}{\electronvolt} for the electronic self-consistency loop and of \SI{1e-3}{\electronvolt\per\angstrom} for the residual forces, respectively.

A thermodynamic model based on the law of mass action was applied to calculate ion speciation and solubility.
This modeling approach is based on existing works \cite{Clark2017, Clark2019, Limpo1993, Limpo1995}.
The cell-level simulations are conducted with a continuum model based on the quasi-particle method derived in our previous works \cite{Clark2017, Clark2019}.
The equilibrium calculations from the thermodynamic model are integrated into the cell-level simulations, assuming that complex formation reactions are much faster than typical time scales of the charge and discharge.
The model consisted of a system of 12 equations: 4 electrolyte-conservation-equations describing the electrolyte speciation, 3 solid-volume-conservation equations, 3 solute mass continuity equations, electrolyte-charge continuity expression and 1 expression representing either the galvanostatic or potentiostatic condition.
A P2D finite-volume model, with spatial resolution in electrolyte transport and cathodic diffusion, was implemented in Python.
The differential-algebraic equations were solved with MATLABs fully-implicit ode15s solver. 

The cell model was parametrized based on recent designs for \textdelta-\ce{MnO2} 2032-like coin cells as presented in literature \cite{Alfaruqi2015c,Ren2019,Chen2022,Chen2022a}.
Parameters are mostly taken from the coin cells manufactured in the recent study of Chen and coworkers \cite{Chen2022}, which are similar to most other designs.
Cathode composition is a mixture of \ce{MnO2}, acetylene black and a PVDF binder with 70:20:10 \si{wt\percent} with a mass loading of \SI{2}{\milli\gram\per\cm\squared}.
Relative volume fractions are calculated based on the theoretical densities of the materials.
Pore volume measurements were reported in the studies from Shen et. al. \cite{Shen2021} and Corpuz et. al. \cite{Corpuz2019} in the range of \SIrange{0.44}{0.78}{\cm\cubic\per\gram}.
Here, we use a pore volume of \SI{0.5}{\cm\cubic\per\gram} to calculate the porosity of the cathode and, combined with the mass loading, the resulting cathode thickness of \SI{66}{\micro\meter}.
The separator thickness is set to \SI{150}{\micro\meter} \cite{Qin2020}.
If not stated otherwise, the electrolyte used is an aqueous solution of \SI{2}{\Molar} \ce{ZnSO4}, \SI{0.5}{\Molar} \ce{MnSO4} and cycling of the cell is simulated under galvanostatic conditions at \SI{200}{\milli\ampere\per\gram}.
The full details of the calculation and choice of parameters can be found in the Supporting Information.  

\FloatBarrier
\medskip


\medskip
\textbf{Acknowledgments} \par 

The authors acknowledge support from the Helmholtz Association, the state of Baden-Wuerttemberg through bwHPC, and the German Research Foundation (DFG) through Grant No. INST 40/467-1 FUGG (JUSTUS cluster).
Part of this work was performed on the HoreKa supercomputer funded by the Ministry of Science, Research and the Arts Baden-Wuerttemberg and by the Federal Ministry of Education and Research.
The research leading to these results has received funding from the Federal Ministry of Education and Research (BMBF) in the framework of the project 'ZIB' (FKZ 03XP0204A).
Further support by the German Research Foundation (DFG) under Germany's Excellence Strategy - EXC 2154 - Project number 390874152 is gratefully acknowledged.

\medskip


\medskip

\iftoggle{Wiley}
{
	\pagebreak
	
	\begin{figure}
		\textbf{Table of Contents}\\
		\medskip
		\includegraphics[width=55mm]{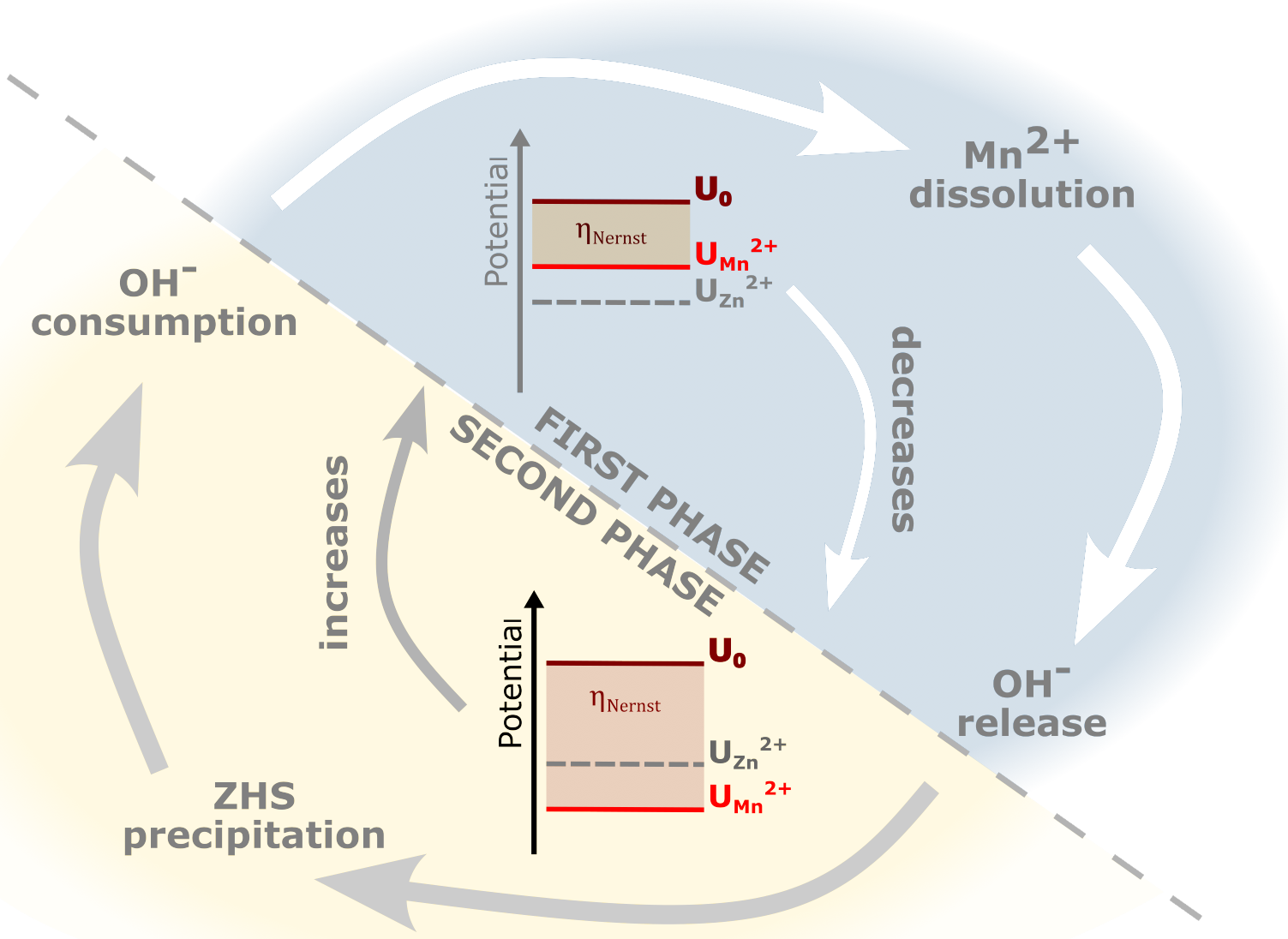}
		\medskip
		\caption*{The two-phase cycling behavior of \ce{MnO2}-based ZIB in near-neutral aqueous electrolytes 
			is experimentally well documented, however, the cycling mechanism remains disputed. Our theory-based full-cell model describes the dynamics resulting from the interplay of electrolyte precipitation and cathodic dissolution. This mechanism is validated with experimental evidence and used to optimize cycling behavior.}
	\end{figure}
}
{}

\bibliographystyle{apsrev4-2}
\bibliography{bibliography}	

\end{document}